\documentclass[11pt]{article}
\usepackage{lipsum} 
\usepackage{datetime}

\usepackage[utf8]{inputenc}
\usepackage{url}
\usepackage{setspace}
\usepackage{palatino}
\usepackage{graphicx}
\usepackage{float}
\usepackage{titling} % drop vertical space before the title
\usepackage{multirow}
\usepackage{lscape}
\usepackage{amsmath}
\usepackage{amssymb}
\usepackage{xspace}
\usepackage{subcaption}
\usepackage[a4paper, total={6in, 9.5in}]{geometry}
\fontfamily{ppl}\selectfont 
\usepackage[american]{babel}
\usepackage[
pdfa=true
]{hyperref}

\title{Rare decays of the SM Higgs to light pseudoscalars in the CMS Experiment}
\author{Presented at the 32nd International Symposium on \\ Lepton Photon Interactions at High Energies,\\
Madison, Wisconsin, USA\\\\\\
Anagha Aravind \\ University of Wisconsin-Madison \\ {\it On behalf of the CMS Collaboration}
}    
\date{August 25-29, 2025}

\begin{document}

\maketitle
\selectlanguage{american}
\begin{abstract}
An overview of selected searches where the SM Higgs boson decays to a pair of light pseudoscalars is presented using Run 2 CMS data at $\sqrt{s}=13$ TeV. Final states with b quarks, leptons, and photons are discussed.
\end{abstract}

\selectlanguage{american} 

\section{Introduction}
A recent CMS measurement sets a limit of 12\% at 95\% CL on the decay of the Standard Model (SM) Higgs boson to beyond SM particles~\cite{CMS:detector,CMS:portraitHiggs2022}. The SM Higgs boson has a narrow decay width $\simeq 4.07$ MeV, which implies that even exotic processes with small branching fractions become relevant for new physics searches~\cite{Curtin:2013ExoticHiggs}. Interpreting extended Higgs-sector models provides a simple and well-motivated approach to exploring new physics. Such models may address several shortcomings of the SM, including the hierarchy problem, matter–antimatter asymmetry, and the nature of dark matter. 

This article will focus on exotic searches in the CMS detector where the SM Higgs boson decays to two light pseudoscalars, referred to as $a$, with each $a$ boson further decaying to SM particles. One popular theory for the interpretation of these searches is the 2 Higgs Doublet model (2HDM+S)~\cite{theory:2HDMS}. 

\section{Recent $h\rightarrow aa$ results at CMS Experiment}
CMS has a rich program of $h\rightarrow aa$ searches. In this article, selected recent results that analyze the full Run-2 data for $10 \text{ MeV } < m_a< 60$ GeV are discussed.

\subsection{$h \rightarrow aa \rightarrow 4e$}
This analysis probes $h \rightarrow aa$ with $4e$ in the final state, i.e, each $a \rightarrow ee$ where $10 < m_a < 100$ MeV~\cite{CMS:4e}. %Signal samples are generated using POWHEG generator and Higgs related background processes are generated using the MADGRAPH generator. 
The results are interpreted in Axion-Like Particle (ALP) models, where the $a$ pseudoscalars are potential dark matter candidates.

Since the $a$'s have very low masses, the final state $e$'s are highly collimated, and the search is limited by the resolution of the ECAL detector of CMS. Selected events are required to have two Merged Electron Pairs (MEPs), which are reconstructed from pairs of tracks in the silicon tracker matched to ECAL super clusters within a distance of $\Delta R <0.015$. In addition, an Adaptive Vertex Fitter is used to fit the vertex of the selected MEPs. Isolation cuts are applied to the selected MEPs to further suppress the background from QCD activity.

Major backgrounds in the analysis are SM photon conversions, SM processes with electrons and QCD multijet. To optimize signal extraction and background suppression, a multivariate identification technique is used, taking as input various characteristics of the MEPs. 

\begin{figure}[h!]
        \centering
        \includegraphics[width=0.35\textwidth]{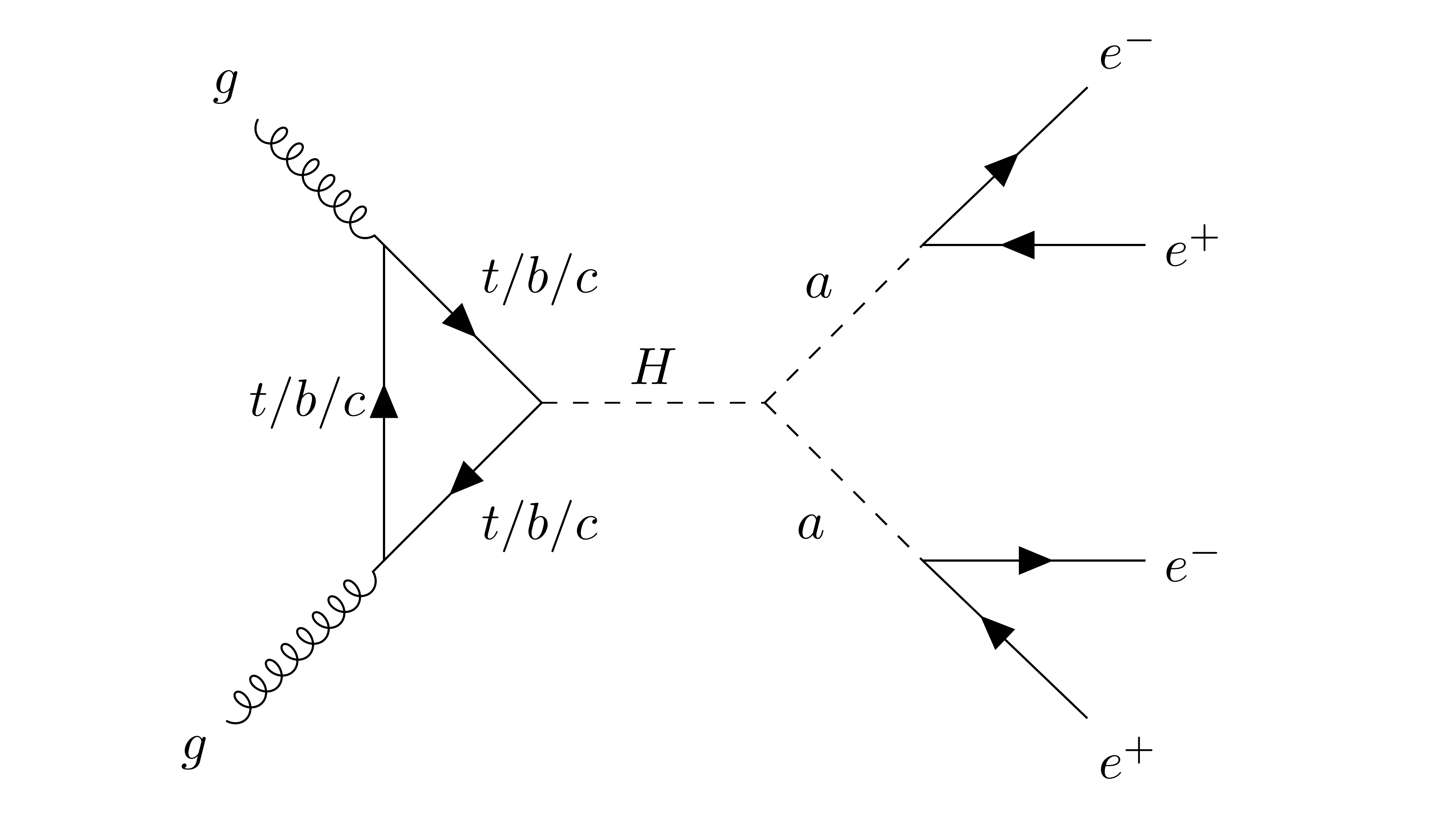}
        \includegraphics[width=0.30\textwidth]{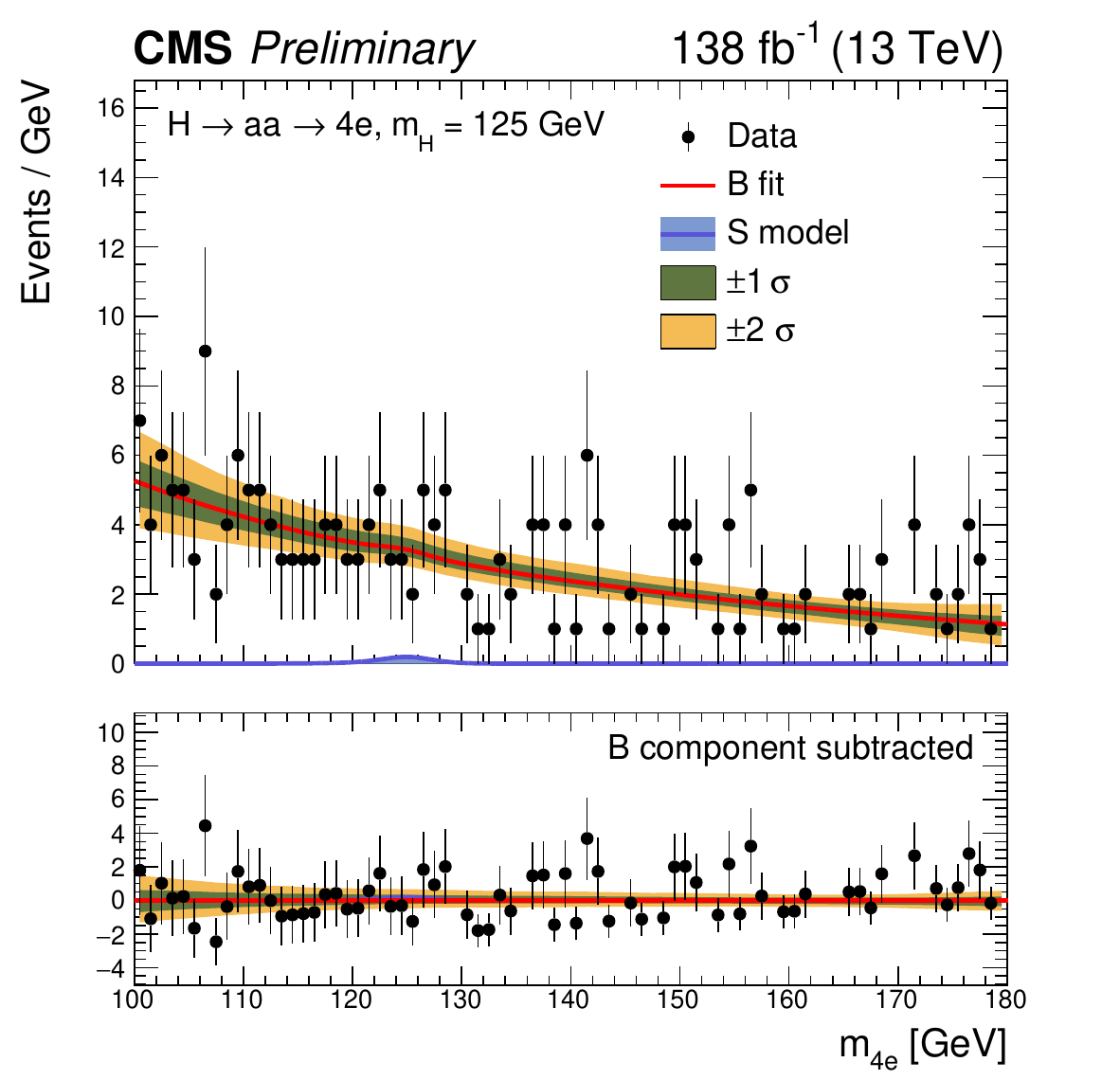}
        \includegraphics[width=0.30\textwidth]{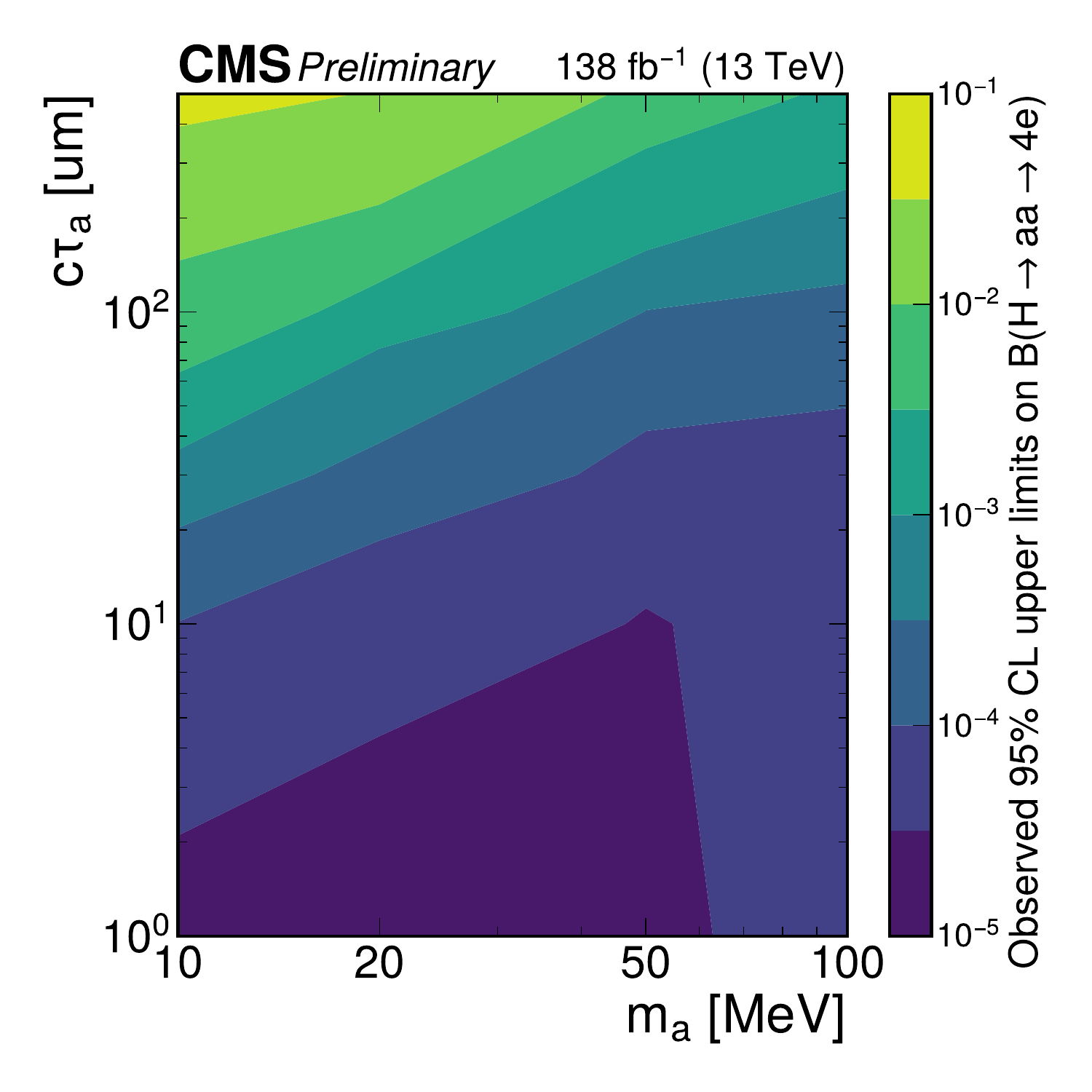}
        \caption{Left: Feynmann diagram of $h \rightarrow aa \rightarrow 4e$ where the Higgs is produced by gluon-gluon fusion. Middle: Invariant mass distribution of the 2 MEPs ($4e$) system compared to a fit of the background (red) and signal (blue) for $m_a= 20$ GeV and $c\tau = 10\mu m$. Right: The observed limits at 95\% CL on the B($h \rightarrow aa \rightarrow 4e$) in the 2D decay length vs $m_a$ plane~\cite{CMS:4e}.}
        \label{fig:4e}
\end{figure}

The signal is extracted from the reconstructed Higgs boson mass of the two MEP system by fitting the data to a set of parametric shapes depending on the physics process. Stringent limits are set on the B($h \rightarrow aa \rightarrow 4e$) for different lifetimes of the ALP, with lower lifetimes being the most sensitive as shown in Fig.~\ref{fig:4e}.
 
\subsection{$h \rightarrow aa \rightarrow 4\gamma$}
A search is performed for $h \rightarrow aa \rightarrow 4\gamma$, where one pair of $\gamma\gamma$ is reconstructed as fully resolved photons and the other pair is reconstructed as a merged photon~\cite{CMS:4gamma}. The analysis is interpreted in NMSSM, ALP, and 2HDM theories. %Signal samples are generated with MADGRAPH5\_aMC@NLO, and processed through the CMS detector simulation based on GEANT4. 

Events are required to pass diphoton triggers. Selected events require exactly 3 reconstructed $\gamma$, where the invariant mass of the $\gamma$, $m_{\gamma\gamma\gamma} >90$ GeV. The analysis strategy is to extract the $h \rightarrow aa \rightarrow 4\gamma$ signal from the 2D $m_{a}$ spectra in the signal region defined by $\Delta m_a \leq 2 $GeV and $100<m_{\gamma\gamma\gamma}<140$ GeV. The $a$ decaying to resolved $\gamma\gamma$ is reconstructed from the four-momentum sum of the resolved photons. To reconstruct the merged $\gamma\gamma$ pair, a Graphical Neural Network (GNN) is used, with energy deposit maps from the ECAL as input. 

\begin{figure}[h!]
        \centering
        \includegraphics[width=0.50\textwidth]{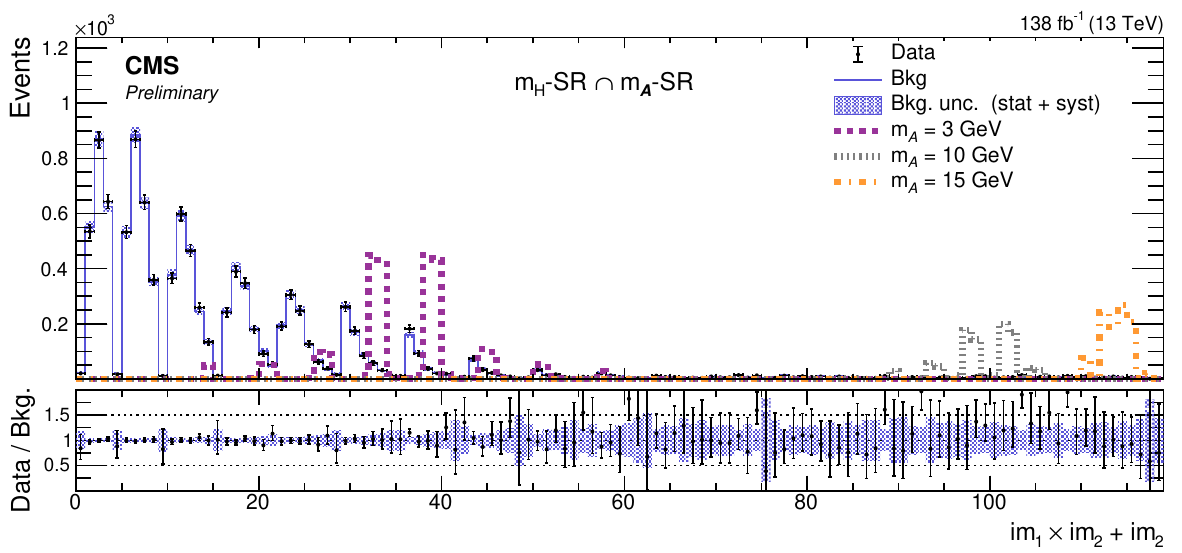}
        \includegraphics[width=0.35\textwidth]{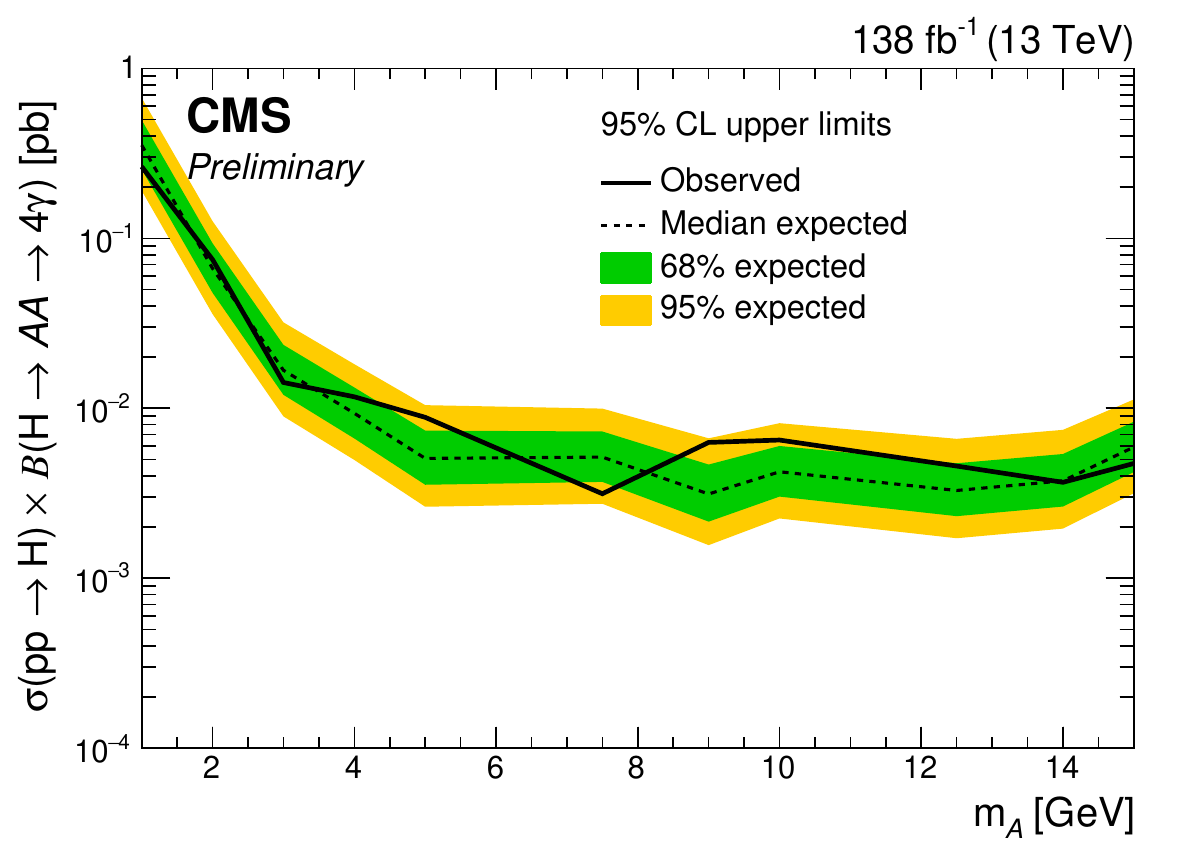}
        
        \caption{Left: Unrolled 2D-$m_a$ spectra in the final signal region obtained by fixing $m_{A,1}$, while scanning along increasing bins of $m_{A,2}$. Right: Observed and median expected upper limits at 95\% CL for $h \rightarrow aa \rightarrow 4\gamma$~\cite{CMS:4gamma}.}
        \label{fig:hto2ph}
\end{figure}

A side-band region is used for estimating the dominant backgrounds, such as QCD multijet, photon+jet and prompt diphoton, from data. This analysis sets an upper limit at 95\% CL on the B($h \rightarrow aa \rightarrow 4\gamma$) of 0.264 to 0.008 pb for $1 < m_a <15$ GeV as shown in Fig.~\ref{fig:hto2ph}.

\subsection{$h \rightarrow aa \rightarrow 4\tau / 2\mu2\tau$}
This analysis probes $h \rightarrow aa$ decay, where one $a$ decays to $\tau\tau$ and the other $a$ decays to either $\tau\tau$ or $\mu\mu$~\cite{CMS:4tau}. The final state is $(\tau_{\mu}\tau_{\text{1-prong}})(\tau_{\mu}\tau_{\text{1-prong}})$, i.e., for each $a$, one $\tau$ decays into a $\mu$ and the other is a leptonic or 1-prong hadronic $\tau$ decay. Since the light $a$ bosons are Lorentz boosted, the final products are highly collimated. 
%The signal and background samples are generated using a combination of PYTHIA, MADGRAPH\_aMC@NLO and POWHEGBOX, and GEANT4 is used for CMS detector simulation.

Events must pass dimuon trigger. Selected events require two same-signed muons that are well-separated, i.e. $\Delta R >1.5$ and a charged track in the vicinity of each of these muons. Depending on the isolation and kinematics, these tracks are classified into "isolation" and "signal" (with tighter cuts) tracks. The "isolation" tracks are used for background estimation and "signal" tracks are used in signal region. The signal is extracted from a 2D map filled with ordered pairs of ($m_1$,$m_2$) where $m_2>m_1$. 

The major background is QCD multijet, with minor contributions from $t\bar{t}$, W+Jets, Z+Jets, and diboson processes. These are modeled in the 2D distribution from 1-D distributions of the $\mu$-track system transformed by a matrix dependent on the correlation between the 2 reconstructed $a$ bosons.

\begin{figure}[h!]
        \centering
        \includegraphics[width=0.30\textwidth]{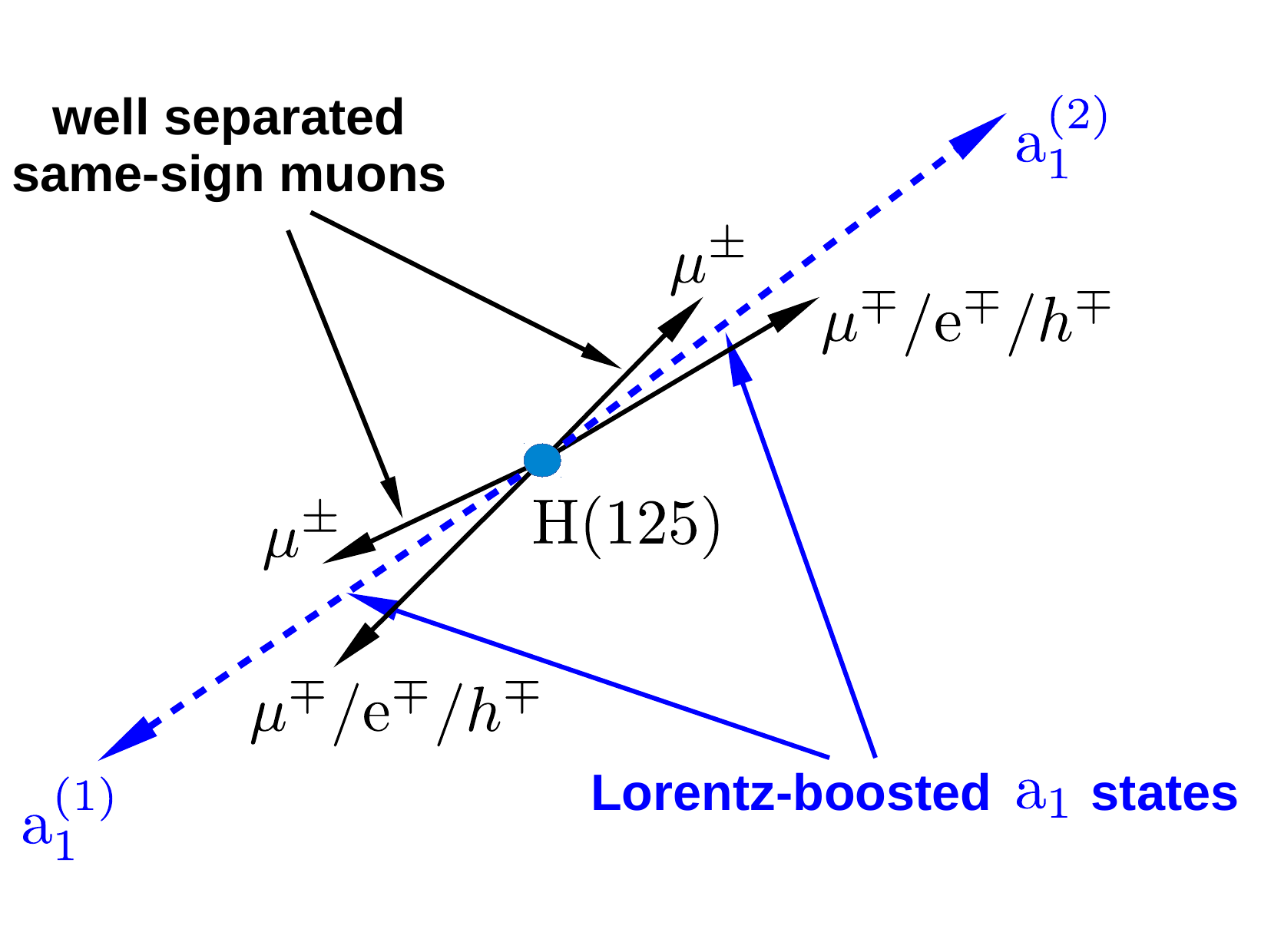}
        \includegraphics[width=0.27\textwidth]{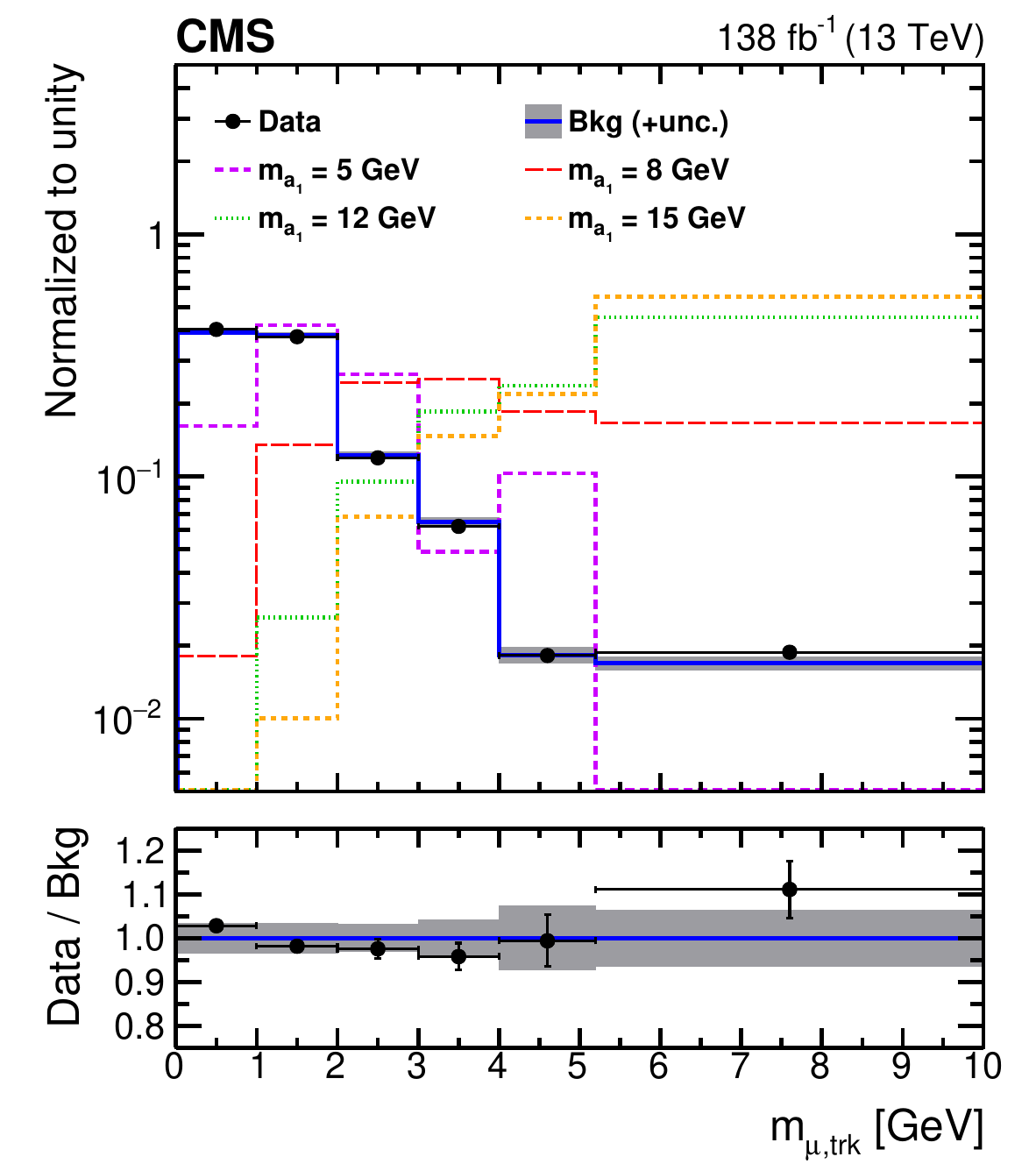}
        \includegraphics[width=0.30\textwidth]{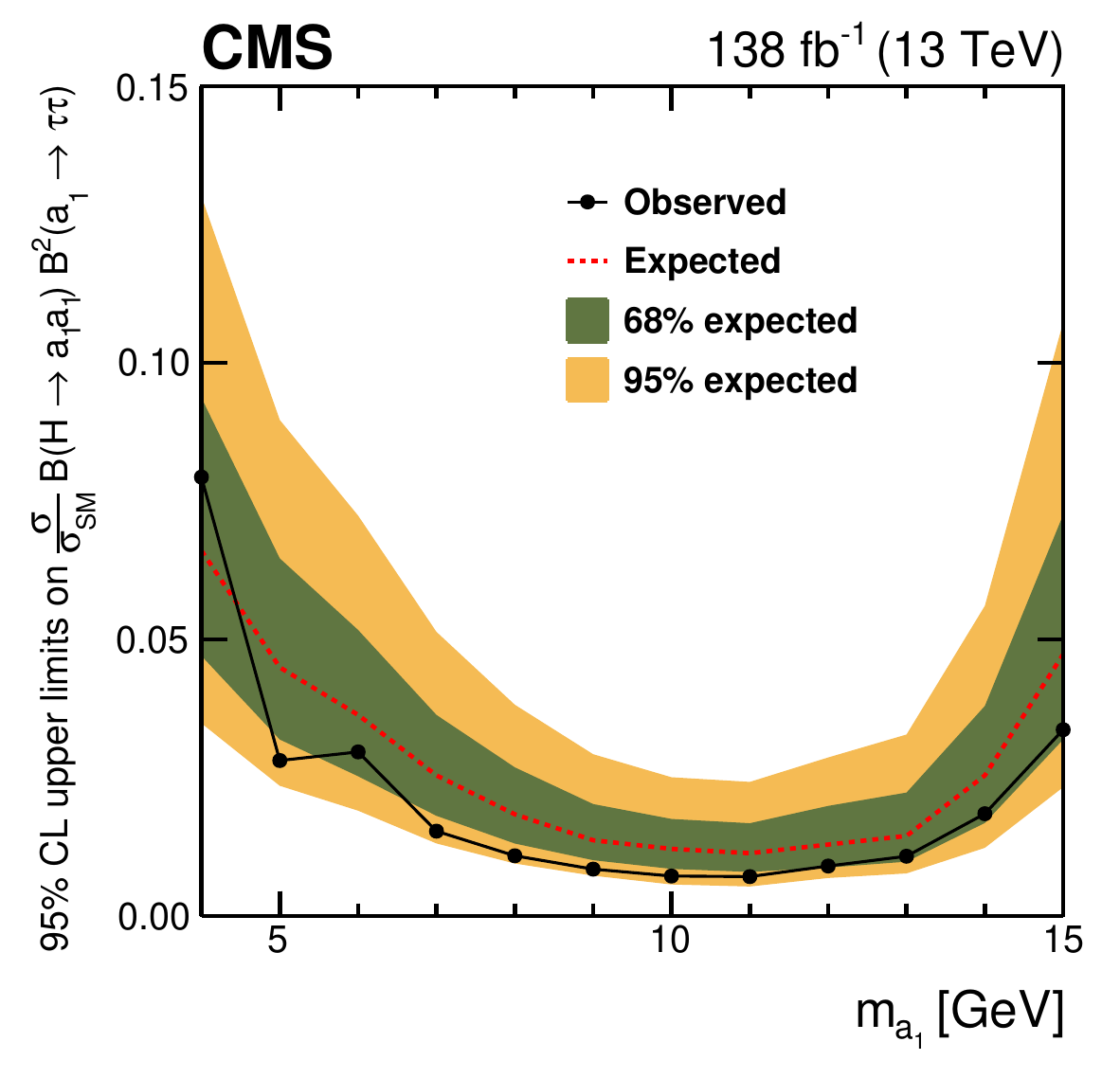}
        \caption{Left: Signal topology for $h \rightarrow aa \rightarrow 4\tau / 2\mu2\tau$. Middle: Normalized invariant mass distribution of the $\mu$-track system for events passing signal selection. Right: Observed and median expected upper limits at 95\% CL for $\sigma(pp \rightarrow H+ X ) B(h \rightarrow aa)B(a\rightarrow \tau\tau)$ relative to the SM Higgs boson production cross section ($\sigma_{SM}$)~\cite{CMS:4tau}.}
        \label{fig:4tau}
\end{figure}

The model independent observed limits at 95\% CL on the $\sigma(pp \rightarrow H+ X ) B(h \rightarrow aa)B(a\rightarrow \tau\tau)$ relative to the SM Higgs boson production cross section ($\sigma_{SM}$) ranges from 0.007 to 0.079 for $4\leq m_a \leq15$ GeV. Most stringent limits are set in the Type III of the 2HDM+S interpretation as shown in Fig.~\ref{fig:4tau}.

\subsection{$h \rightarrow aa \rightarrow bb\tau \tau$}
A search is performed for $h \rightarrow aa \rightarrow bb\tau \tau$ where $12< m_a< 60$ GeV~\cite{CMS:bbtt}. The $\tau$s decay further and hence, three channels are considered: $e\mu$, $e\tau_h$, and $\mu \tau_h$. The $bb\tau\tau$ final state is chosen for it's favourable predicted branching fraction when interpreted in the 2HDM+S theory. 

Selected events must pass either single lepton or di-lepton triggers depending on the channel. Each event must have exactly one $\tau\tau$ pair and at least one b-tagged jet. The two legs of the $\tau\tau$ pair must have opposite charges and a $\Delta R > 0.4$. The major backgrounds are QCD multijet background, jets misidentified as $\tau_h$ fakes, $Z\rightarrow \tau \tau$, $t\bar{t}$ and are estimated from a combination of data-driven methods and simulation.

Deep Neural Networks (DNN) are trained for each channel for the purpose of signal optimization, taking advantage of several distinguishing kinematic features between signal and background. The signal regions are defined by cutting on the transformed DNN score. The signal is then extracted from the invariant mass distribution of the reconstructed $\tau \tau$ system in the signal region using a maximum likelihood fit method. 

\begin{figure}[h!]
        \centering
        \includegraphics[width=0.32\textwidth]{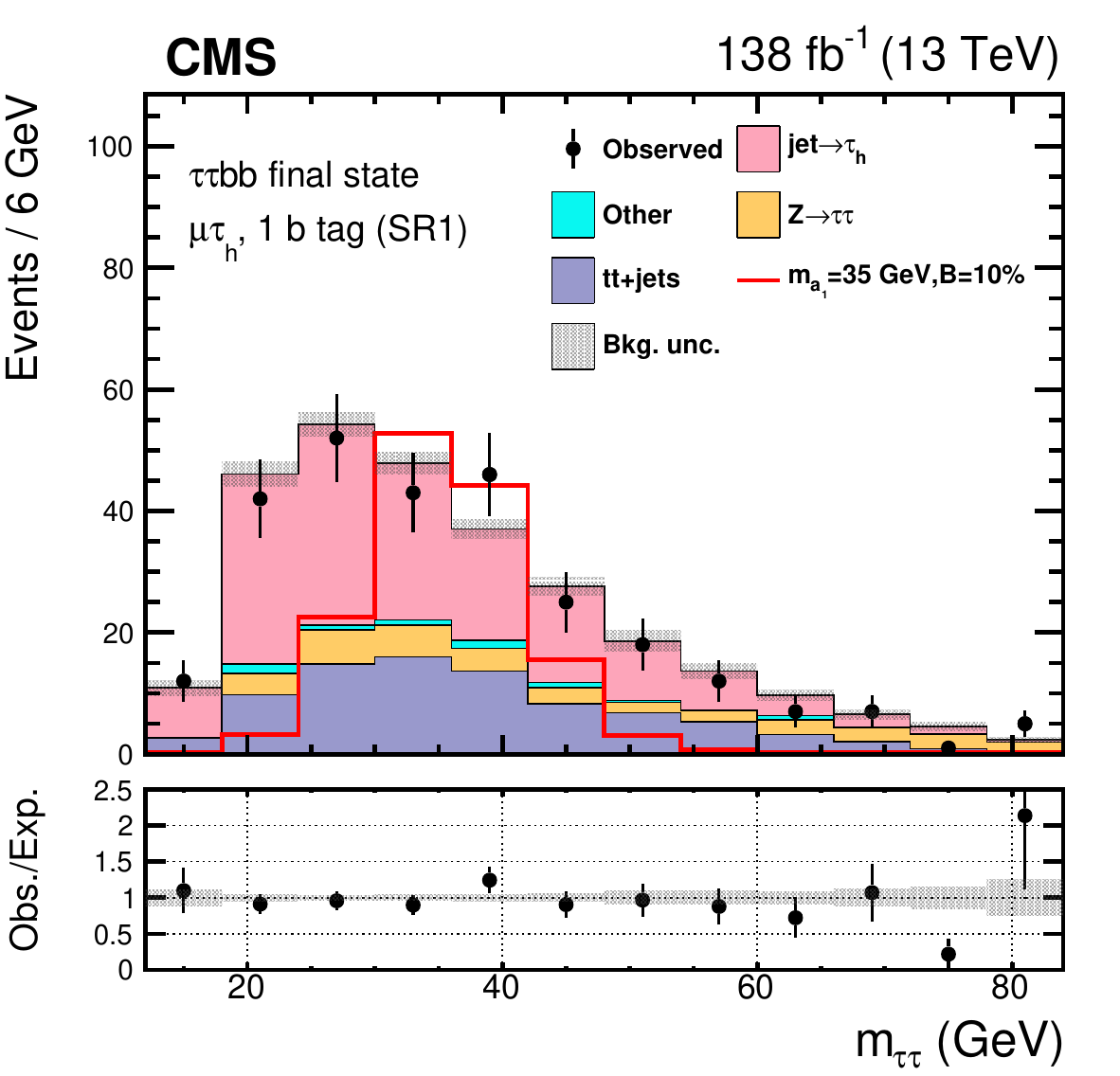}
        \includegraphics[width=0.28\textwidth]{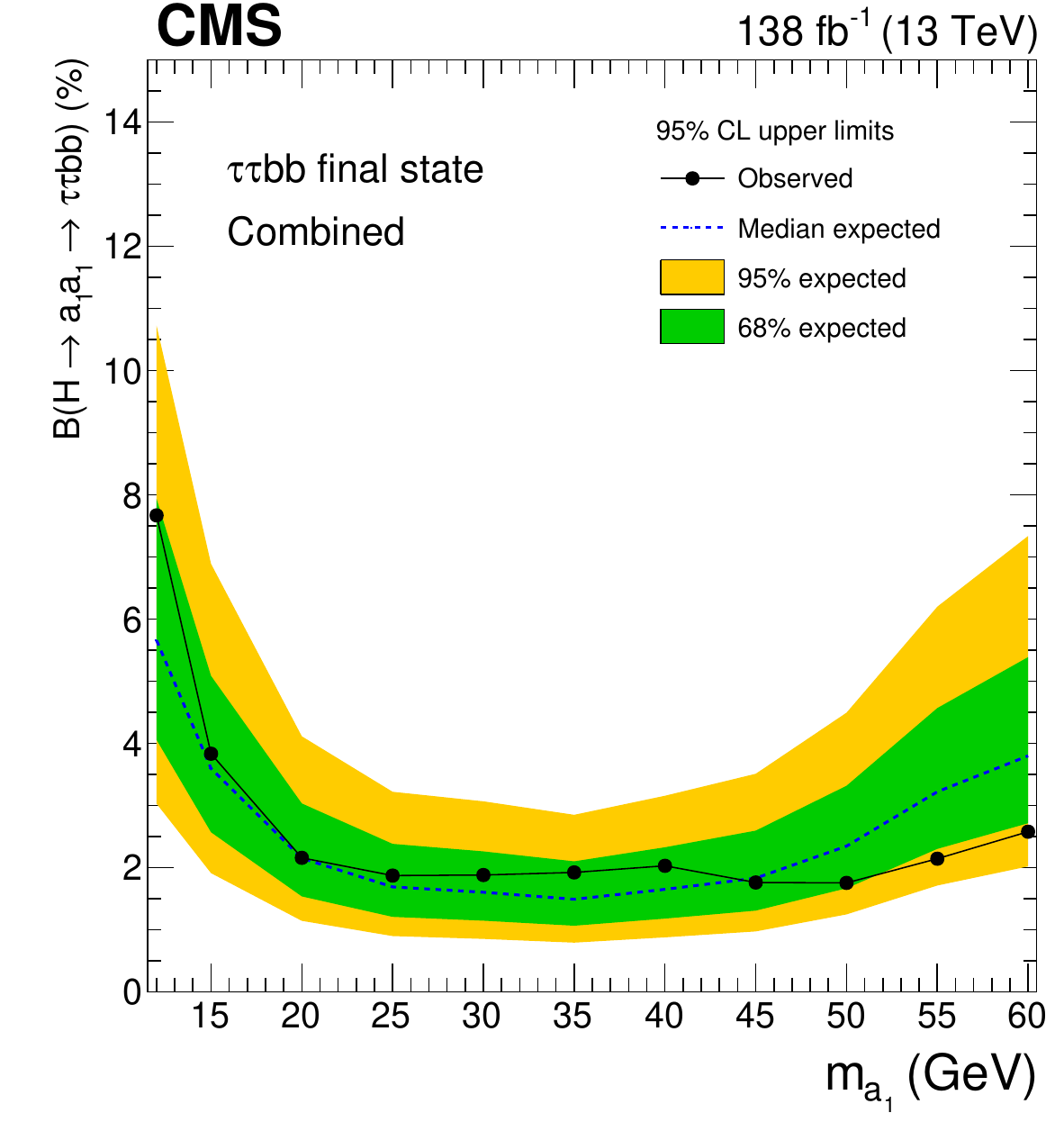}
        \includegraphics[width=0.30\textwidth]{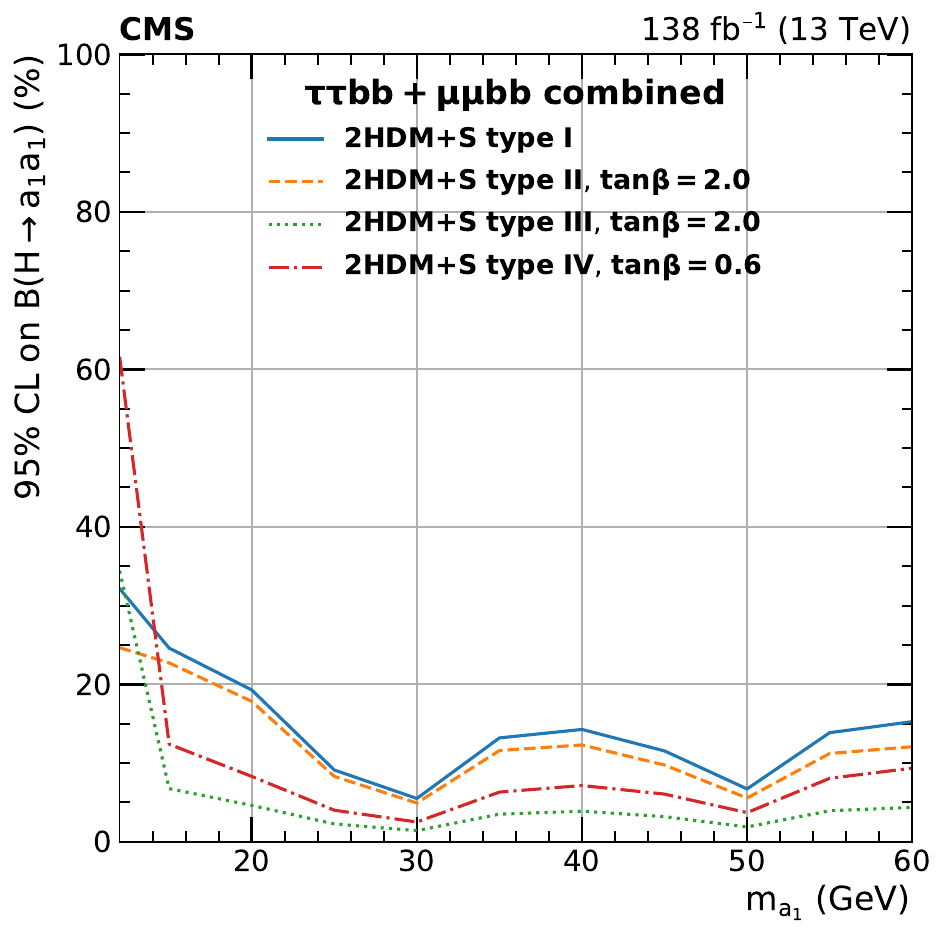}
        \caption{Left: $m_{\tau\tau}$ distribution for $\mu\tau$ channel in one of the signal regions. Middle: Observed and median expected upper limits at 95\% CL for $h \rightarrow aa \rightarrow 2b2\tau$ for the combination of all 3 channels: $e\mu$, $e\tau_h$ and $\mu \tau_h$. Right: Observed upper limits at 95\% CL for $h \rightarrow aa$ obtained from the combination of $2b2\tau$ and $\mu\mu bb$ final states for 2HDM+S Type I (independent of $\tan b$), Type II ($\tan b=2.0\%$), Type III ($\tan b=2.0$), and Type IV ($\tan b=0.6$), respectively~\cite{CMS:bbtt}.}
        \label{fig:2b2t}
\end{figure}

This analysis sets an upper limit at 95\% CL on the B($h \rightarrow aa \rightarrow bb\tau \tau$) at 1.8--7.7\% for $12< m_a< 60$ GeV for observed data. The results are interpreted in the 2HDM+S theory and also shown in combination with the $bb\mu\mu$ final state as shown in Fig.~\ref{fig:2b2t}. 

\section{Conclusion}The $h \rightarrow aa \rightarrow 4e$ result is the first $h\rightarrow aa$ search in CMS to set limits for  $10 < m_a < 100$ MeV. The analysis with $h \rightarrow aa \rightarrow 4\gamma$ bridges the gap between two previous analyses in CMS (one with fully merged and the other with fully resolved final state~\cite{CMS:2023HAA4gammaMerged, CMS:2023HAA4gammaResolved}). The $h \rightarrow aa \rightarrow 4\tau / 2\mu2\tau$ result shows improvements over the previous CMS result~\cite{CMS:20164tau} by a factor of 2. Similarly, $h \rightarrow aa \rightarrow bb\tau\tau$ in combination with $bb\mu\mu$ sets stringent limits for B($h \rightarrow aa \rightarrow bbll$) in the 2HDM+S model interpretation, for 12--60 GeV. 

While several $h\rightarrow aa$ searches with different final states have been performed in CMS, no significant deviation from the SM have been observed. However, there is potential for other final states to be explored in the future, including searches with asymmetric decays, i.e. $h\rightarrow a_1 a_2$ where $a_1, a_2$ have different masses.

\begin{figure}[h!]
        \centering
        \includegraphics[width=0.34\textwidth]{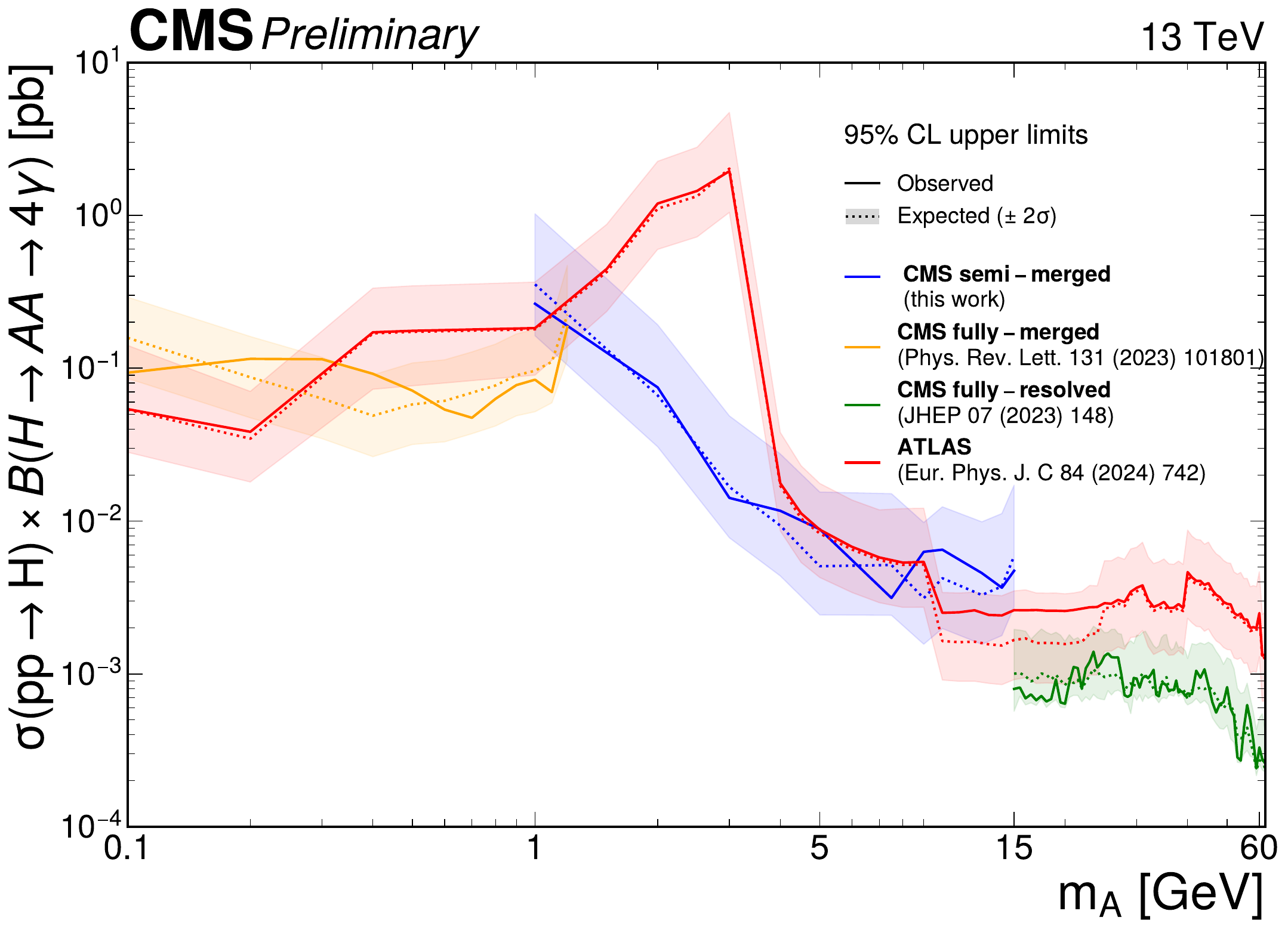}
        \includegraphics[width=0.32\textwidth]{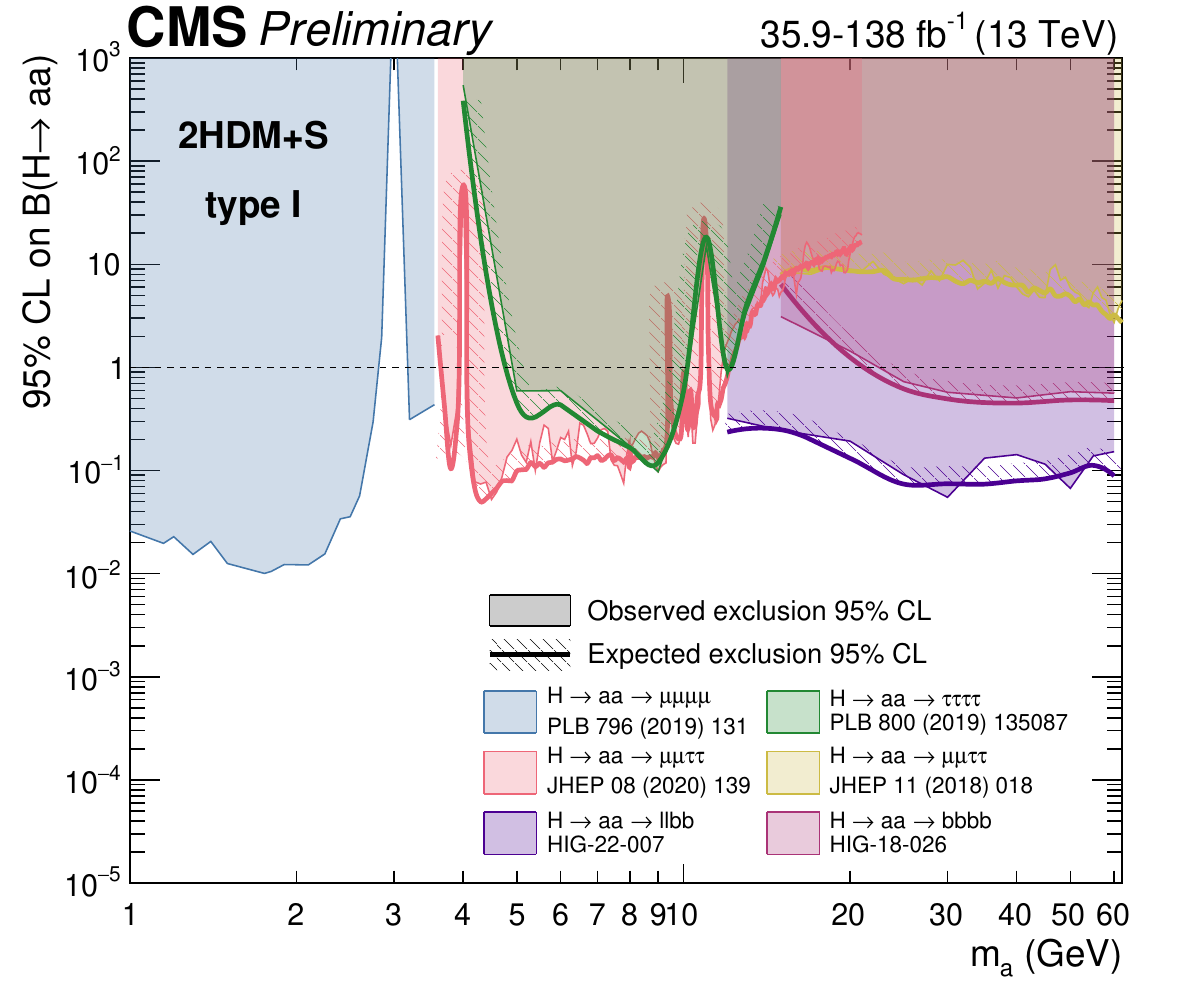}
        \caption{Left: CMS results for $h \rightarrow aa \rightarrow 4\gamma\gamma$ in different mass ranges shown in comparison with results from the ATLAS experiment~\cite{CMS:4gamma, CMS:2023HAA4gammaMerged, CMS:2023HAA4gammaResolved}. Right: CMS summary plot of recent  $h\rightarrow aa$ results for 2HDM+S Type I interpretation~\cite{CMS:summary2HDMSRun2}.}
        \label{fig:summary-plots}
\end{figure}

\bibliographystyle{unsrt}
\bibliography{LPreferences}

@article{CMS:portraitHiggs2022,
  author = {CMS Collaboration},
  title         = {A portrait of the {Higgs} boson by the {CMS} experiment ten years after the discovery},
  journal       = {Nature},
  volume        = {607},
  number        = {7917},
  pages         = {60--68},
  year          = {2022},
  url={\url{http://dx.doi.org/10.1038/s41586-022-04892-x}},
  note={doi: \url{10.1038/s41586-022-04892-x}. URL \url{http://dx.doi.org/10.1038/s41586-022-04892-x}},
}

@article{CMS:summary2HDMSRun2,
  author = {CMS Collaboration},
  title         = {Summary of {2HDM+S} searches at 13 {TeV (Run 2)}},
  journal       = {CMS Public TWiki},
  year          = {2025},
  note          = {\url{https://twiki.cern.ch/twiki/bin/view/CMSPublic/Summary2HDMSRun2}}
}

@article{Curtin:2013ExoticHiggs,
  title = {Exotic decays of the 125 {GeV} {Higgs} boson},
  author = {Curtin, David and others},
  journal = {Phys. Rev. D},
  volume = {90},
  issue = {7},
  pages = {075004},
  numpages = {91},
  year = {2014},
  month = {Oct},
  publisher = {American Physical Society},
  note = {doi: \url{10.1103/PhysRevD.90.075004}. URL \url{https://link.aps.org/doi/10.1103/PhysRevD.90.075004}},
}

@article{theory:2HDMS,
title = {Theory and phenomenology of two-{Higgs}-doublet models},
journal = {Physics Reports},
volume = {516},
number = {1},
pages = {1-102},
year = {2012},
issn = {0370-1573},
note = {doi: \url{10.1016/j.physrep.2012.02.002}. URL \url{https://www.sciencedirect.com/science/article/pii/S0370157312000695}},
author = {G.C. Branco and others},
}

@article{CMS:4e,
      author = {CMS Collaboration},
      collaboration = "CMS",
      title         = "{Search for light pseudoscalar a bosons in $\mathrm{H\to a
                       a \to 4 e}$ decays in pp collisions at $\sqrt{s}$ = 13
                       TeV}",
      institution   = "CERN",
      reportNumber  = "CMS-PAS-EXO-24-031",
      journal = {CMS Physics Analysis Summary CMS-PAS-EXO-24-031},
      address       = "Geneva",
      year          = "2025",
      note           = {URL \url{https://cds.cern.ch/record/2932415}},
}

@article{CMS:4gamma,
      collaboration = "CMS",
      author = {CMS Collaboration},
      title         = "{Search for exotic Higgs boson decays $H \to
                       \mathcal{A}\mathcal{A}$ with $\mathcal{A} \to \gamma\gamma$
                       in events with a partially merged topology in proton-proton
                       collisions at $\sqrt{s} = 13~\mathrm{TeV}$}",
      institution   = "CERN",
      reportNumber  = "CMS-PAS-EXO-24-025",
      journal = {CMS Physics Analysis Summary CMS-PAS-EXO-24-025},
      address       = "Geneva",
      year          = "2025",
      note            = {URL \url{https://cds.cern.ch/record/2937956}},
}

@article{CMS:4tau,
  author = {CMS Collaboration},
  title         = "{Search for light pseudoscalar boson pairs produced from
                       Higgs boson decays using the 4$ \tau $ and 2$ \mu2\tau $
                       final states in proton-proton collisions at $ \sqrt{s}= $
                       13 {TeV}}",
    institution   = "CERN",
    archivePrefix = "arXiv",
    eprint        = "2508.06947",
    reportNumber  = "CMS-SUS-24-002, CERN-EP-2025-129, CMS-SUS-24-002-003",
    journal       = "Submitted to JHEP",
    address       = "Geneva",
    year          = "2025",
    note            = {URL \url{https://arxiv.org/abs/2508.06947}},
}

@article{CMS:2023HAA4gammaMerged,
  author = {CMS Collaboration},
  collaboration = "CMS",
      title         = "{Search for exotic Higgs boson decays H $ \to \mathcal{A}
                       \mathcal{A} \to 4\gamma$ with events containing two merged
                       diphotons in proton-proton collisions at $ \sqrt{s} = $ 13
                       TeV}",
      archivePrefix = "arXiv",
      eprint        = "2209.06197",
      reportNumber  = "CMS-HIG-21-016, CERN-EP-2022-151, CMS-HIG-21-016-004",
      journal       = "Phys. Rev. Lett.",
      volume        = "131",
      pages         = "101801",
      year          = "2023",
      note          = {doi: \url{10.1103/PhysRevLett.131.101801}. URL \url{https://cds.cern.ch/record/2826997}},
}

@article{CMS:2023HAA4gammaResolved,
  author = {CMS Collaboration},
  title         = "{Search for the exotic decay of the Higgs boson into two
                       light pseudoscalars with four photons in the final state in
                       proton-proton collisions at $ \sqrt{s} $ = 13 TeV}",
      archivePrefix = "arXiv",
      eprint        = "2208.01469",
      reportNumber  = "CMS-HIG-21-003, CERN-EP-2022-095, CMS-HIG-21-003-003",
      journal       = "JHEP",
      volume        = "2307",
      pages         = "148",
      year          = "2023",
      note          = {doi: \url{10.1007/JHEP07(2023)148} URL \url{https://cds.cern.ch/record/2823906}},      
}

@article{CMS:20164tau,
  author = {CMS Collaboration},
  title         = "{Search for light pseudoscalar boson pairs produced from
                       decays of the 125 GeV Higgs boson in final states with two
                       muons and two nearby tracks in pp collisions at $\sqrt{s} =
                       $ 13 TeV}",
      archivePrefix = "arXiv",
      eprint        = "1907.07235",
      reportNumber  = "CMS-HIG-18-006, CERN-EP-2019-105, CMS-HIG-18-006-003",
      journal       = "Phys. Lett. B",
      volume        = "800",
      pages         = "135087",
      year          = "2020",      
      note           = {doi: \url{10.1016/j.physletb.2019.135087}. URL  \url{https://cds.cern.ch/record/2682977}},
}

@article{CMS:bbtt,
  author = {CMS Collaboration},
title         = "{Search for exotic decays of the Higgs boson to a pair of
                       pseudoscalars in the $ \mu\mu\mathrm{b}\mathrm{b} $ and $
                       \tau\tau\mathrm{b}\mathrm{b} $ final states}",
      archivePrefix = "arXiv",
      eprint        = "2402.13358",
      reportNumber  = "CMS-HIG-22-007, CERN-EP-2023-284, CMS-HIG-22-007-003",
      journal       = "Eur. Phys. J. C",
      volume        = "84",
      pages         = "493",
      year          = "2024",
      note           = {doi: \url{10.1140/epjc/s10052-024-12727-4}. URL \url{https://cds.cern.ch/record/2889907}},
}

@article{CMS:detector,
note = { doi: \url{10.1088/1748-0221/3/08/S08004}. URL  \url{https://doi.org/10.1088/1748-0221/3/08/S08004}},
year = {2008},
month = {Aug},
publisher = {},
volume = {3},
number = {08},
pages = {S08004},
author = {CMS Collaboration},
title = {The {CMS} experiment at the {CERN LHC}},
journal = {Journal of Instrumentation},
}
\end{document}